\documentclass[10pt,aps,prb,showpacs,superscriptaddress,floatfix,
twocolumn]{revtex4}
\usepackage{amsmath}
\usepackage{citesort}
\usepackage{graphicx}
\begin{document}
\title{On the Relation between Optical Conductivity and Quasiparticle
Dynamics: Boson Structures}
\author{J.P. Carbotte}
\affiliation{Department of Physics and Astronomy, McMaster University,\\
Hamilton, Ontario, L8S 4M1 Canada}
\author{E. Schachinger}
\email{schachinger@itp.tu-graz.ac.at}
\homepage{www.itp.tu-graz.ac.at/~ewald}
\affiliation{Institute of Theoretical and Computational Physics\\
Graz University of Technology, A-8010 Graz, Austria}
\author{J. Hwang}
\affiliation{Department of Physics and Astronomy, McMaster University,\\
Hamilton, Ontario, L8S 4M1 Canada}
\date{\today}
\begin{abstract}
An extended Drude form is often used to analyze optical data in
terms of an optical scattering rate and renormalized mass
corresponding, respectively, to the real and imaginary part of
the memory function. We study the relationship between memory
function and quasiparticle self energy for an isotropic system.
We emphasize particularly boson signatures.
We find it useful to introduce
a new auxiliary model scattering rate and its Kramers-Kronig
transform determined solely from optics which are much
closer to the self energy than is the memory function itself
in the normal state. In the superconducting state the simplification
fails because the quasiparticle density of states acquires an
essential energy dependence.
\end{abstract}
\maketitle
\newpage
\section{Introduction}

Optical\cite{tanner,puchkov} and angular resolved photo emission
(ARPES) data\cite{campuzano,damascelli,norman} have given a
wealth of information on quasiparticle
dynamics in the cuprates both in their normal and superconducting
state. The methods are complementary but the exact quantitative
relationship between the two is complex. One important difference
is that ARPES gives direct information on angular variations
around the Fermi surface while optics involves an average over all
the quasiparticles participating in the absorption.
Even if these anisotropies are not accounted for (isotropic
system) there remain additional differences which have their
fundamental origin in the fact that ARPES measures directly
the quasiparticle spectral density $A({\bf k},\omega)$ at
fixed quasiparticle momentum {\bf k} as a function of energy
$\omega$ while optics involves the current-current correlation
function which depends on the product of two spectral
densities\cite{mars1,schach1} at the same momentum {\bf k}
but with frequencies displaced by the photon energy.

In this paper we focus on similarities as well as on essential
differences between the information that is derived from these
two different probes and particularly on how they are to be
compared. We will be interested in both, the normal and the
superconducting state at zero and at finite temperature. In the
literature on optics it is the temperature and frequency dependence
of the optical scattering rate $\tau^{-1}_{op}(T,\omega)$ which
has been particularly emphasized.\cite{puchkov} More recently
the imaginary part\cite{hwang} of the memory function\cite{quijada,%
allen} related to $\tau^{-1}_{op}(T,\omega)$ by Kramers-Kronig (KK)
transform has also been used to compare directly with the
energy dependence of the real part of the quasiparticle self
energy determined by ARPES.\cite{lanzara,zhou} Of particular
interest in such a comparison is the relationship between the
position in frequency of peak structures seen in these quantities
and how they reflect corresponding peaks in the electron-boson
spectral density $I^2\chi(\omega)$.\cite{mars2,verga,schach2}
For phonons $I^2\chi(\omega)$ is the well known
electron-phonon spectral density while in the cuprates exchange of
spin fluctuations\cite{schach1,schach3,schach4,schach5} is
more appropriate although as yet there exists no consensus%
\cite{chubukov} on this issue. Here we will consider several models
for $I^2\chi(\omega)$ within an extended Eliashberg formalism. In
the superconducting state provision is made for $d$-wave symmetry
of the gap. To keep things simple the same form of $I^2\chi(\omega)$
is used in both gap and renormalization channels but with
different magnitudes with the ratio between the two equal to $g$.

In Sec.~\ref{sec:2} we consider only the normal state and coupling of
quasiparticles to a single $(\omega_E)$ Einstein mode.\cite{mars1,%
carb1} We also employ a simplified approximate scattering time
formulation of the optical conductivity formula in the normal state.
This allows some analytic results to be established and provides
the motivation for introducing a new model scattering rate and its
KK-transform determined solely from optical data but which is
very close to the self energy itself. While the real part of the
self energy has a logarithmic singularity at $\omega_E$ the
imaginary part of the memory function does not. Instead, it
shows a peak at $\omega=\sqrt{2}\omega_E$. On the other hand,
our model quantities reproduce well the self energy at $T=0$
and only very small differences arise at finite $T$. In Sec.~\ref{sec:3}
we provide the more general formalism needed to treat the
conductivity accurately as well as to take care of extended
$I^2\chi(\omega)$ spectra and the superconducting state. Results
in these cases are presented in Sec.~\ref{sec:4}.
For the superconducting state
additional complications arise because of the essential energy
dependence of the self consistent quasiparticle density of states
(DOS). In Sec.~\ref{sec:5} we present our numerical results for our new
model scattering rate based on the complete formula for the
optical conductivity and solutions of the Eliashberg equations
and compare these with the imaginary part of the quasiparticle
self energy. We confirm the previous expectation, based on
simplifying assumptions, that in the normal state these two
quantities are close in magnitude as well as frequency dependence.
In the superconducting state they are not. Finally, conclusions
are found in Sec.~\ref{sec:6}.

\section{Normal State and Coupling to a Single Boson}
\label{sec:2}

A generalized Drude form with frequency dependent optical scattering
rate $\tau^{-1}_{op}(\omega)$ and optical effective mass ratio
$[m^\ast(\omega)/m]_{op}$ has been used extensively to analyze
optical data in a one component approach. The optical conductivity
$\sigma(\omega)$ is written in terms of the memory function\cite{quijada}
$M(\omega) = \tau^{-1}_{op}(\omega)-i\omega\lambda_{op}(\omega)$
with $\lambda_{op}(\omega) = \{[m^\ast(\omega)/m]_{op}-1\}$ the optical
mass renormalization parameter. We have\cite{zhou}
\begin{equation}
  \label{eq:1}
  \sigma(\omega) = \sigma_1(\omega)+i\sigma_2(\omega) = 
  \frac{\Omega_p^2}{4\pi}\frac{i}
  {i\tau^{-1}_{op}(\omega)+\omega\left[1+\lambda_{op}(\omega)\right]}
\end{equation}
and $\tau^{-1}_{op}(\omega)$ and $i\omega\lambda_{op}(\omega)$ are
related by Kramers Kroning, so that when the optical scattering rate
is known, the optical mass renormalization can be obtained from
the KK-transform.

The conductivity is a two particle property given by the
current-current correlation function. It is related in a complicated
way to the quasiparticle self energy (a one particle property)
$\Sigma(\omega)$. In the normal state\cite{mars2} at zero temperature
$(T=0)$
\begin{equation}
  \label{eq:2}
  \sigma(\omega) = \frac{\Omega_p^2}{4\pi}\frac{i}{\omega}
  \int\limits_0^\omega\!d\nu\,\frac{1}
  {\omega+i\tau^{-1}_{imp}-\Sigma(\nu)-\Sigma(\omega-\nu)},
\end{equation}
where $\tau^{-1}_{imp}$ is the impurity scattering rate and
$\Sigma(\nu)$ accounts for the inelastic scattering. In terms of
the electron-phonon or electron-spin fluctuation spectral density
$\alpha^2F(\omega)$ or $I^2\chi(\omega)$ respectively\cite{mars3,%
grimvall} [call it $F(\omega)$]
\begin{eqnarray}
  \Sigma(\omega) &=& \Sigma_1(\omega)+i\Sigma_2(\omega) =
  \int\limits_0^\infty\!d\Omega\,F(\Omega)
  \ln\left\vert\frac{\Omega-\omega}{\Omega+\omega}\right\vert\nonumber\\
 &&-i\pi\int\limits_0^{\vert\omega\vert}\!d\Omega\,F(\Omega).
  \label{eq:3}
\end{eqnarray}
For coupling to a single Einstein mode $F(\omega) = A\delta(\omega-
\omega_E)$ we obtain the simple, well known formula
\begin{equation}
  \label{eq:4}
  \Sigma(\omega) = A\ln\left\vert\frac{\omega_E-\omega}
  {\omega_E+\omega}\right\vert-i\pi A\theta(\vert\omega\vert-\omega_E),
\end{equation}
with $\theta(x)$ the Heavyside $\theta$-function. It is clear from
\eqref{eq:4} that the imaginary part has a finite jump at $\omega_E$
as well as a logarithmic singularity in its real part which can be
used to identify boson structure in both cases. The self energy
can in principle be determined from ARPES data which measures the
quasiparticle spectral function
\begin{equation}
  \label{eq:5}
  A({\bf k},\omega) = -\frac{1}{\pi}{\rm Im}G({\bf k},\omega+i0^+),
\end{equation}
with the Greens function $G({\bf k},\omega+i0^+) = [\omega-
\varepsilon_{\bf k}-\Sigma(\omega+i0^+)]^{-1}$, where
$\varepsilon_{\bf k}$ is the quasiparticle dispersion.
A usefull approximate expression for $\tau^{-1}_{op}(\omega)$ was
obtained directly in second order perturbation theory by
P.B.~Allen.\cite{allen1} It is (for $T=0$)
\begin{equation}
  \label{eq:6}
  \tau^{-1}_{op}(\omega)\simeq\frac{2\pi}{\omega}
  \int\limits_0^\omega\!dz\,F(z)(\omega-z)+\tau^{-1}_{imp}.
\end{equation}
Details in the validity of this form, based on numerical evaluation
of Eqs.~\eqref{eq:2} and \eqref{eq:3} are found in Refs.~\onlinecite{mars2}
and \onlinecite{shulga}.
For $\omega>0$
\begin{equation}
  \label{eq:7}
   \tau^{-1}_{op}(\omega) = \frac{2\pi A_{tr}}{\omega}\left(
   \omega-\omega_E\right)\theta\left(\omega-\omega_E\right),
\end{equation}
where the subscript $tr$ means transport, although for simplicity,
here we will not make a distinction between quasiparticle and
transport spectral density; in general they are different.
(We have set the impurity term equal to zero for simplicity.) 
Note that this formula is less favorable for the identification of
boson structures than is the imaginary part of the self energy.
While it is zero for $\omega <\omega_E$ and finite for
$\omega\ge\omega_E$ it has no jump at $\omega_E$. Instead, it
smoothly increases from its zero value which makes it harder to
identify the exact position of $\omega_E$ in $\tau^{-1}_{op}(\omega)$.
\begin{figure}[t]
\vspace*{-6mm}
  \includegraphics[width=9cm]{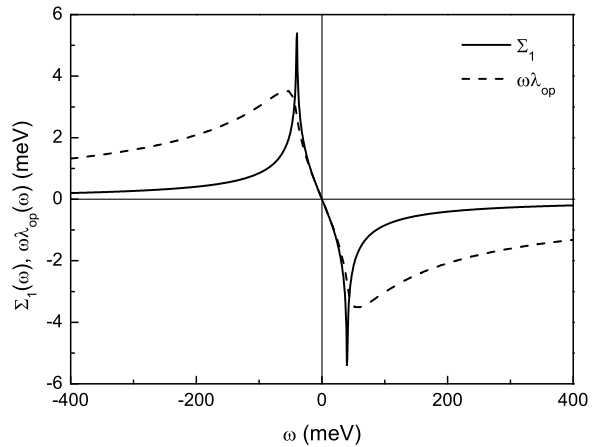}
\vspace*{-6mm}
  \caption{Comparison of the real part of the quasiparticle self energy
$\Sigma_1(\omega)$ as a function of $\omega$ with the corresponding
optical quantity $\omega\{[m^\ast(\omega)/m]_{op}-1\}\equiv\omega%
\lambda_{op}(\omega)$. Coupling is to a single Einstein oscillator with
$\omega_E=40\,$meV and $A=A_{tr}=1\,$meV in Eqs.~\protect{\eqref{eq:4}} and
\protect{\eqref{eq:8}}. There is a logarithmic singularity in
$\Sigma_1(\omega)$ at $\omega=\omega_E$ while $\omega\lambda_{op}%
(\omega)$ only has a small peak at $\omega=\sqrt{2}\omega_E$ and
a logarithmic singularity in the slope at $\omega=\omega_E$.
}
  \label{fig:1}
\end{figure}
Application of the KK-transform to Eq.~\eqref{eq:7} gives
immediately
\begin{equation}
  \label{eq:8}
  \omega\lambda_{op}(\omega) = -2A\left(\ln\left\vert\frac{\omega_E+\omega}
  {\omega_E-\omega}\right\vert+\frac{\omega_E}{\omega}\ln
  \left\vert\frac{\omega_E^2-\omega^2}{\omega_E^2}\right\vert
  \right)
\end{equation}
as the optical mass enhancement. This is to be compared with the
real part of the quasiparticle self energy, Eq.~\eqref{eq:4}. In
terms of the usual mass enhancement parameter $\lambda$
[defined as two times the first inverse moment of $F(\omega)$],
$A=\lambda\omega_E/2$, and the limit of $\Sigma_1(\omega)$ as $\omega\to 0$
equals $-\omega\lambda$. Thus the slope of the self energy
gives the value of $\lambda$ as does also $\omega\lambda_{op}\sim\lambda%
\omega$. This is seen in
Fig.~\ref{fig:1} where we compare $\Sigma_1(\omega)$
(solid line) and $\omega\lambda_{op}(\omega)$ vs $\omega$ (dashed line)
for parameters $A = 1\,$meV and $\omega_E = 40\,$meV. While these
functions agree in the $\omega\to 0$ limit they deviate from each
other at finite frequencies. In particular, at $\omega = \omega_E$
the quasiparticle self energy exhibits a logarithmic singularity while
the optical mass renormalization $\omega\lambda_{op}(\omega)$ has only a
logarithmic singularity in its slope. It can, furthermore, be
shown that $\omega\lambda_{op}(\omega)$ has its maximum at $\omega=\sqrt{2}%
\omega_E$. Thus, while the boson structure in $\Sigma(\omega)$
appears at the boson energy and is singular, the maximum in
$\lambda_{op}(\omega)$ is instead displaced to $\sqrt{2}\omega_E$ and
is associated with a rather broad peak in comparison. We conclude
from these considerations that the optical mass enhancement parameter
$\lambda_{op}(\omega)$ can be used to identify boson structure in the
imaginary part of the memory function as done recently by Hwang
{\it et al.}\cite{hwang} but its signature is much weaker than is
the case in the quasiparticle self energy and $\omega_E$ is
displaced by a factor of $\sqrt{2}$, a fact that does not appear
to have been appreciated before, but is important to realize.

In preparation for what will come later when we consider the
superconducting state, we note that the self energy \eqref{eq:4}
and, consequently, the corresponding optical quantity \eqref{eq:8}
would also apply\cite{li,mitro1,mitro2} in an impurity model
with constant quasiparticle scattering rate $\tau^{-1}_{qp} = A$,
but in which the normal state quasiparticle density of states,
$N(\varepsilon)$, has a gap of size $\omega_E$ at the Fermi
surface, i.e.: $N(\varepsilon) = N_0$ for $\varepsilon<0$ and
$\varepsilon>\omega_E$ but is zero for $0<\varepsilon<\omega_E$.
This shows immediately that structure in $N(\varepsilon)$ can have
an effect on the self energy which is, in some cases,
indistinguishable from boson structure. We return to this important
point later.

\begin{figure}[t]
  \includegraphics[width=9cm]{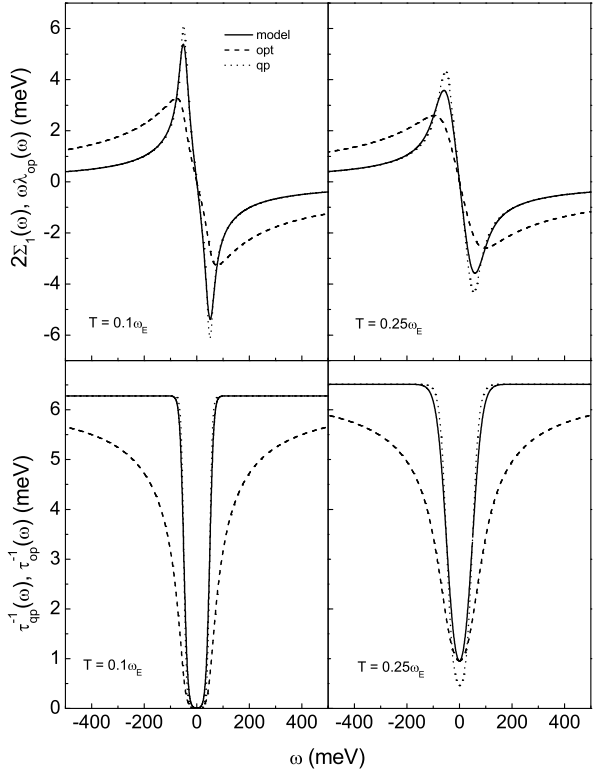}
  \caption{Top frames give a comparison of the real part of the
quasiparticle self energy $2\Sigma_1(\omega)$
(dotted line) with its optical (dashed line) and
model (solid line) counterparts as a function of $\omega$. The right hand
frame is for temperature $T=0.25\omega_E$ and the left hand
frame is for $T=0.1\omega_E$. The bottom frames gives
corresponding quasiparticle scattering rates on which the results
in the two top frames are based. The parameters are $A=1\,$meV and
$\omega_E=40\,$meV.
}
  \label{fig:2}
\end{figure}

How is this picture changed when we consider finite temperatures?
In this case the formulas determining the conductivity as well as the
quasiparticle self energy are more complex. Nevertheless, a simple
picture still emerges. The normal state optical conductivity is
now given by\cite{lee}
\begin{eqnarray}
  \sigma(\omega) &=& \frac{\Omega_p^2}{4\pi}\frac{1}{i\omega}
  \left\{\int_{-\infty}^0\!d\nu\,\tanh\left(\frac{\nu+\omega}{2T}\right)
  S^{-1}(T,\omega,\nu)\nonumber\right.\\
  &&+\left.\int_0^\infty\!d\nu\,\left[\tanh\left(\frac{\nu+\omega}{2T}
  \right)-\tanh\left(\frac{\nu}{2T}\right)\right]\right.\nonumber\\
 &&\times S^{-1}(T,\omega,\nu)
  \biggr\},
  \label{eq:9}
\end{eqnarray}
where
\begin{equation}
  \label{eq:10}
  S(T,\omega,\nu) = \omega+\Sigma^\ast(T,\nu+\omega)-
  \Sigma(T,\nu)-\tau^{-1}_{imp},
\end{equation}
and the self energy is\cite{mars3}
\begin{eqnarray}
  \Sigma(T,\omega) &=& -\int\!dz\,F(z)\left[\psi\left(\frac{1}{2}+i
  \frac{\omega+z}{2\pi T}\right)\right.\nonumber\\
  &&\left.-\psi\left(\frac{1}{2}+
  i\frac{\omega-z}{2\pi T}\right)\right],
  \label{eq:11}
\end{eqnarray}
where $\psi(z)$ is the digamma function and $\Sigma^\ast(T,\omega)$ is
the complex conjugate of $\Sigma(T,\omega)$. For a general $F(z)$
\begin{eqnarray}
  \tau^{-1}_{op}(T,\omega) &=& -2\Sigma_2(T,\omega)\nonumber\\
  &=& \pi\int\!dz\,F(z)\left[
  2\coth\left(\frac{z}{2T}\right)\right.\nonumber\\
  &&\left.-\tanh\left(\frac{\omega+z}{2T}\right)
  +\tanh\left(\frac{\omega-z}{2T}\right)\right].\nonumber\\
  \label{eq:12}
\end{eqnarray}
Shulga {\it et al.}\cite{shulga} were able to show that to a good
approximation
\begin{eqnarray}
  \tau^{-1}_{op}(T,\omega)&\simeq&\frac{\pi}{\omega}\int\!dz\,F(z)
  \left[2\omega\coth\left(\frac{z}{2T}\right)\right.\nonumber\\
  &&\left.-(\omega+z)
  \coth\left(\frac{\omega+z}{2T}\right)\right.\nonumber\\
  &&\left.+(\omega-z)\coth\left(
  \frac{\omega-z}{2T}\right)\right].
  \label{eq:13}
\end{eqnarray}
At zero temperature Eqs.~\eqref{eq:12} and \eqref{eq:13} reduce
to the imaginary part of Eq. \eqref{eq:3} and to Eq.~\eqref{eq:6},
respectively. KK-transforms can be used to get from Eqs.~\eqref{eq:12}
and \eqref{eq:13} the real part of $\Sigma(T,\omega)$ at finite $T$
as well as $\omega\lambda_{op}(T,\omega)$.
Results are shown in Fig.~\ref{fig:2}
for coupling to a single Einstein mode as in Fig.~\ref{fig:1}. Three
curves are show in each frame. The dotted line applies to the quasiparticle
self energy, the dashed line to the corresponding optical quantity and the
solid line to a model yet to be defined. The top frames give
real parts
at $T=0.1\omega_E$ (left) and at $T=0.25\omega_E$ (right). The
corresponding scattering rates on which these are based, are shown in
the two bottom frames. First we consider the
top frames. In all cases we see that temperature smooths out the
corresponding boson structures, but even for $T=0.25\omega_E$ they
remain easily identifiable, although, the difference between
quasiparticle and optical quantities is no longer as pronounced.

The solid curve in Fig.~\ref{fig:2} is based on the following
observation. At $T=0$, $ d[\omega\tau^{-1}_{op}(\omega)]/d\omega$
given by Eq.~\eqref{eq:6} equals exactly the quasiparticle
scattering rate given as twice the imaginary part of the quasiparticle
self energy in Eq.~\eqref{eq:3}. Thus,
in this particular limit one can get the quasiparticle scattering rate
directly from optics by taking the first derivative of
$\omega\tau^{-1}_{op}(\omega)$. Of course, we have assumed that
anisotropies in momentum space can be neglected. In general, ARPES
will give information on the {\bf k}-dependence of the
quasiparticle self energy while optics is always an average.
Neglecting such complications, it is interesting to introduce a model
scattering rate $\tau^{-1}_{model}\equiv d[\omega\tau^{-1}_{op}(\omega)]%
/d\omega$ defined from optics alone and consider its relationship to
the quasiparticle self energy at finite temperatures. The formula for
$\tau^{-1}_{model}(T,\omega)$ equivalent to Eqs.~\eqref{eq:12} and
\eqref{eq:13} is
\begin{eqnarray}
  \tau^{-1}_{model}(T,\omega) &=& \pi\int\!dz\,F(z)\left[
  2\coth\left(\frac{z}{2T}\right)\right.\nonumber\\
  &&\left.+\frac{1-4\left(\frac{\omega-z}{2T}
  \right)}
  {\sinh^2\left(\frac{\omega-z}{2T}\right)}-
  \frac{1-4\left(\frac{\omega+z}{2T}\right)}
  {\sinh^2\left(\frac{\omega+z}{2T}\right)}
  \right].
  \label{eq:14}
\end{eqnarray}

In all cases considered in Fig.~\ref{fig:2} the model quantity (solid
lines) deviates from its quasiparticle counter part (dotted lines) only very
slightly while it deviates much more from the optical case (dashed lines).
 It is
clear, therefore, that if one wishes to compare optical and
quasiparticle quantities  directly, it is better to use the
KK-transform of $\tau^{-1}_{model}(\omega)$ than that for
$\tau^{-1}_{op}(\omega)$ itself. The peaks in the corresponding
model mass enhancement will now correspond to the Einstein
frequency $\omega_E$ and it will be sharper than in the optical
quantity itself which, as we have seen, is also shifted by $\sqrt{2}$.

Note also, that on comparing top and bottom frame it is the function
$\tau^{-1}_{op}(\omega)$ which is least useful in localizing boson
structures as discussed in relation to Eq.~\eqref{eq:7}. All the
other functions have a sharper signature of the Einstein frequency
$\omega_E$.

We next turn to a discussion of how the use of an extended spectrum
for $F(\omega)$ modifies our simple results and how the superconducting
condensation further modifies boson structures in quasiparticle and
optical quantities.

\section{Extended Boson Spectrum and Superconducting State}
\label{sec:3}

In general, coupling of the electrons to a boson spectrum such as
phonons or spin fluctuations is not restricted to a single mode. The
corresponding $\alpha^2F(\omega)$ (phonons) or $I^2\chi(\omega)$
(spin fluctuations) can extend to $100\,$meV or so and even up to
$400\,$meV (of the order of $J$ in the $t-J$ model\cite{sorella}),
respectively.
Here we wish to consider such extended spectra and also consider the
superconducting state assuming $d$-wave gap symmetry since we have
the cuprates in mind. The equation for $\sigma(\omega)$\cite{carb2}
\begin{subequations}
\label{eq:16}
\begin{equation}
  \sigma(T,\nu) = \frac{\Omega^2_p}{4\pi}\frac{i}{\nu}
     \left\langle
      \int\limits_0^\infty\!d\omega\,\tanh\left(\frac{\beta\omega}{2}
        \right)\left[
        J(\omega,\nu)- J(-\omega,\nu)
      \right]\right\rangle_\theta,  \label{eq:16a}
\end{equation}
where $\langle\cdots\rangle_\theta$ denotes the averaging over angle
$\theta$ and the function $J(\omega,\nu)$ is given by
\begin{eqnarray}
  2J(\omega,\nu) &=& \frac{1}{E(\omega;\theta)
   +E(\omega+\nu;\theta)}\left[1-N(\omega;\theta)\right.\nonumber\\
  &&\left.\times N(\omega+\nu;\theta)
  -P(\omega;\theta)P(\omega+\nu;\theta)\right]\nonumber\\
  &&+\frac{1}{E^\ast(\omega;\theta)-
    E(\omega+\nu;\theta)}\nonumber\\
  &&\times\left[1+N^\ast(\omega;\theta)N(\omega+\nu;\theta)\right.
   \nonumber\\
  &&\left.  +P^\ast(\omega;\theta)P(\omega+\nu;\theta)\right],
  \label{eq:16b}
\end{eqnarray}
with
\begin{equation}
  E(\omega;\theta) = \sqrt{\tilde{\omega}^2
   (\omega+i0^+)-\tilde{\Delta}^2
   (\omega+i0^+;\theta)}, \label{eq:16c}
\end{equation}
and
\begin{equation}
  N(\omega;\theta) = \frac{\tilde{\omega}(\omega+i0^+)}
   {E(\omega;\theta)},\qquad
  P(\omega;\theta) = \frac{\tilde{\Delta}(\omega+i0^+;\theta)}
   {E(\omega;\theta)}. \label{eq:16d}
\end{equation}
\end{subequations}
Here $E^\ast$, $N^\ast$, and $P^\ast$ are the complex conjugates
of $E$, $N$, and $P$, respectively.
The two fundamental functions $\tilde{\omega}(\omega)$
and $\tilde{\Delta}(\omega)$ are related the renormalization
and gap functions, respectively.\cite{schach1} They are for the
gap channel
\begin{subequations}
\label{eq:17}
\begin{multline}
  \tilde{\Delta}(\nu+i0^+;\theta) \\
  = \pi Tg
  \sum\limits_{m=0}^\infty\cos(2\theta)\left[\lambda(\nu-i\omega_m)+
  \lambda(\nu+i\omega_m)\right]\\
 \times\left\langle
 \frac{\tilde{\Delta}(i\omega_m;\theta')\cos(2\theta')}{
  \sqrt{\tilde{\omega}^2(i\omega_m)+\tilde{\Delta}^2(i\omega_m;
  \theta')}}\right\rangle_{\theta'}\\
 +i\pi g\int\limits^\infty_{-\infty}\!dz\,\cos(2\theta)
  I^2\chi(z)\left[n(z)+f(z-\nu)\right]\times\\
 \times\left\langle
  \frac{\tilde{\Delta}(\nu-z+i0^+;\theta')\cos(2\theta')}{
  \sqrt{\tilde{\omega}^2(\nu-z+i0^+)-\tilde{\Delta}^2(\nu-z+i0^+;
  \theta')}}\right\rangle_{\theta'},
  \label{eq:17a}
\end{multline}
and in the renormalization channel
\begin{multline}
  \tilde{\omega}(\nu+i\delta)\\
  = \nu+i\pi T
  \sum\limits_{m=0}^\infty\left[\lambda(\nu-i\omega_m)-
  \lambda(\nu+i\omega_m)\right]\\
  \times\left\langle
    \frac{\tilde{\omega}(i\omega_m)}{
  \sqrt{\tilde{\omega}^2(i\omega_m)+\tilde{\Delta}^2(i\omega_m;
  \theta')}}\right\rangle_{\theta'}\\
  +i\pi\int\limits^\infty_{-\infty}\!dz\,
   I^2\chi(z)\left[n(z)+f(z-\nu)\right]\\
  \times\left\langle
  \frac{\tilde{\omega}(\nu-z+i0^+)}{
  \sqrt{\tilde{\omega}^2(\nu-z+i0^+)-\tilde{\Delta}^2(\nu-z+i0^+;
  \theta')}}\right\rangle_{\theta'}
  \label{eq:17b}.
\end{multline}
Here
\begin{equation}
  \label{eq:17c}
  \lambda(\nu) = \int\limits^\infty_{-\infty}\!d\Omega\,
   \frac{I^2\chi(\Omega)}{\nu-\Omega+i0^+}.
\end{equation}
The gap is given by
\begin{equation}
  \label{eq:17f}
  \Delta(\nu+i0^+;\theta) = \nu\,\frac{\tilde{\Delta}(\nu+i0^+;\theta)}{
  \tilde{\omega}(\nu+i0^+)},
\end{equation}
or, if the renormalization function $Z(\nu)$ is introduced in the
usual way as $\tilde{\omega}(\nu+i0^+) = \nu Z(\nu)$ then
\begin{equation}
  \label{eq:17g}
  \Delta(\nu+i0^+;\theta) = \frac{\tilde{\Delta}(\nu+i0^+;\theta)}
  {Z(\nu)}.  
\end{equation}
\end{subequations}
These equations are a minimum set and go beyond a BCS approach.
They include inelastic scattering known to be strong in the
cuprate superconductors. In Eq.~\eqref{eq:17a} $g$ modifies the
electron-boson spectral density $I^2\chi(\omega)$ from its value
in the renormalized channel. In general the shape of
$I^2\chi(\omega)$ could also be different but we have not included
this possible complication here.

In what follows, we will use for $I^2\chi(\omega)$ a form suggested
by Millis {\it et al.}\cite{millis} (MMP) given by
\begin{equation}
  \label{eq:18}
  I^2\chi(\omega) = I^2\frac{\omega/\omega_{SF}}
  {\omega^2+\omega^2_{SF}},
\end{equation}
with $\omega_{SF}$ a spin fluctuation frequency which can be
\begin{figure}[t]
\vspace*{-6mm}
  \includegraphics[width=9cm]{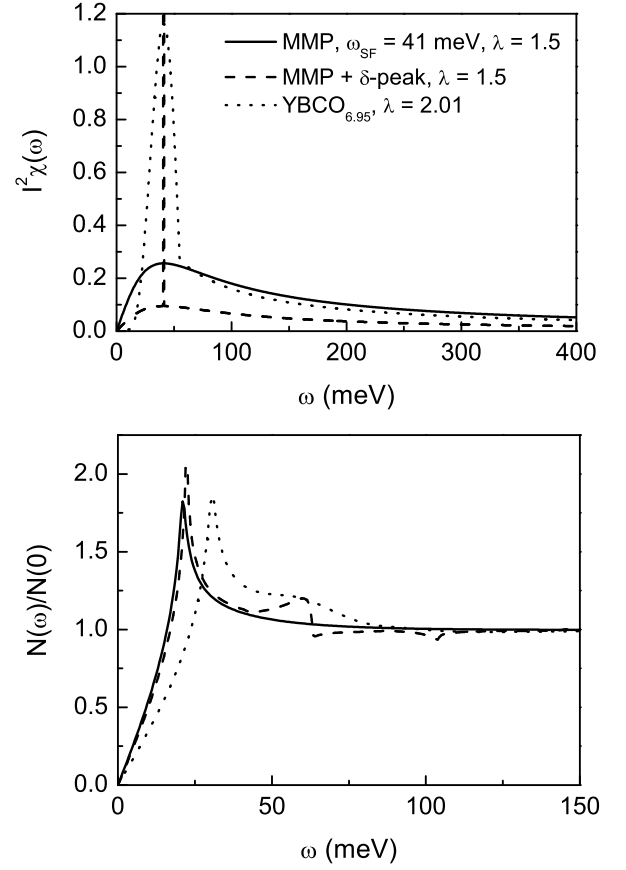}
\vspace*{-4mm}
  \caption{Top frame: the three models for the electron-boson spectral
density $I^2\chi(\omega)$ used in Fig.~\protect{\ref{fig:4}}. The
MMP model with $\omega_{SF}=41\,$meV, $\lambda_{MMP}=0.554$
augmented by a $\delta$-function at $\omega_E=41\,$meV with
$\lambda_\delta=0.946$ for a total of 1.5 (dashed curve). The
model spectrum with $\lambda=2.01$ obtained from a fit to the
optical conductivity as described in Ref.~\onlinecite{schach3}
(dotted curve). Coupling to an optical resonance at $41\,$meV
is included as a peak. A pure MMP spectrum with $\omega_{SF}=41\,$meV
and $\lambda=1.5$ (solid curve). Bottom frame: the superconducting
density of states $N(\omega)/N(0)$ at $T=10\,$K for the three
spectra shown in the top frame.
}
  \label{fig:3}
\end{figure}
fitted to optical data and to emphasize structure we also will add
in one case a $\delta$-function at some specific frequency.
Such an MMP form for $\omega_{SF}=41\,$meV is shown in the top
frame of Fig.~\ref{fig:3} (solid line). It has a $\lambda=1.5$.
Also shown is the same MMP form now with an added $\delta$-function
peak at $\omega_E=41\,$meV (dashed line). $\lambda$ is again equal to
1.5 but now with only 0.554 in the MMP form and the rest in the
$\delta$-function. In the
superconducting state the MMP form is modified because of the
growth of an optical resonance\cite{schach3,schach4,schach5}
at $41\,$meV which is the same frequency at which a spin one
resonance is also observed in spin polarized inelastic neutron
scattering\cite{bourges} for optimally doped YBCO. (In other
cuprates the position of the observed resonance varies somewhat.%
\cite{fong,he}) This optical resonance grows as the temperature
is lowered with the area under the resonance scaling as the
superfluid density\cite{schach1,schach3} to a reasonable
approximation. Such a spectrum has been derived from a fit to
$T=10\,$K optical data on a YBa$_2$Cu$_3$O$_{6.95}$
(YBCO$_{6.95}$) twinned single crystal.\cite{schach3}
This spectrum was also used later on to fit optical
data for untwinned single crystals reported by C.C. Homes
{\it et al.}\cite{schach7,homes} and also to
calculate the microwave conductivity of the YBCO$_{6.99}$
single crystals.\cite{schach6,hoss} This spectrum is shown as
the dotted curve in the top frame of Fig.~\ref{fig:3}.
It has a $\lambda = 2.01$.
We will take this $I^2\chi(\omega)$ spectrum as representative of
the oxides. Note that coupling to a collective mode at $(\pi,\pi)$
is also seen in the ARPES data of Campuzano {\it et al.}\cite{camp1}
Also note that in the above formulation we have neglected possible
momentum space anisotropies. Recent ARPES data by
Kaminski {\it et al.}\cite{kaminski1} justify this assumption.

The bottom frame of Fig.~\ref{fig:3} shows the quasiparticle density
of states $N(\omega)/N(0)$ in the superconducting state for $T=10\,$K
for the three spectra presented in the top frame. This would be the
classic way to see boson structure in conventional superconductors
through tunneling which, so far, has been less successful in
the high $T_c$ oxides. This is the reason why optics and
ARPES became such important tools in studying the quasiparticle
\begin{figure}[tp]
  \includegraphics[width=9cm]{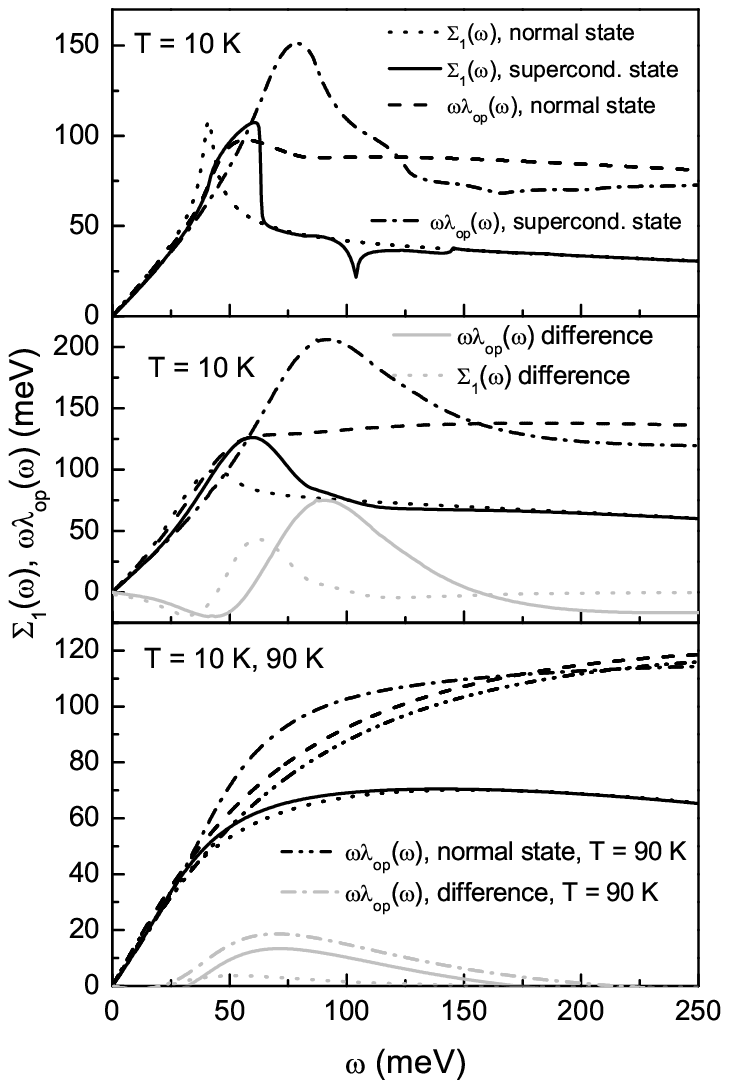}
  \caption{Comparison of the real part of the quasiparticle self
energy $\Sigma_1(\omega)$ vs $\omega$ and optical effective mass
renormalization $\omega\lambda_{op}(\omega)$ for the three model
electron-boson spectral densities $I^2\chi(\omega)$ shown in the
top frame of
Fig.~\protect{\ref{fig:3}}. The top frame is for an MMP form with
$\omega_{SF}=41\,$meV and $\lambda_{MMP} = 0.554$ augmented by
a $\delta$-function at $\omega_E=41\,$meV with $\lambda_\delta=0.946$,
the middle is obtained with a model $I^2\chi(\omega)$ obtained from
consideration of optical properties of YBCO$_{6.95}$
with $\lambda=2.01$ (dotted curve in the top frame of
Fig.~\protect{\ref{fig:3}}),
and the bottom frame is for the pure MMP with $\omega_{SF}=41\,$meV
and $\lambda=1.5$. In all cases the curves come in pairs
$\Sigma_1(\omega)$ in normal (black dotted curve) and superconducting
state (black solid curve) and $\omega\lambda_{op}(\omega)$ in
normal (black dashed curve) and superconducting state (black
dash-dotted curve). In the middle frame we also include as
(gray solid line) the difference between superconducting and
normal state of $\omega\lambda_{op}(\omega)$ and as gray dotted
curve the difference between superconducting and normal state
$\Sigma_1(\omega)$. In the bottom frame similar
difference curves are shown but with two different values of $T$ in the
normal state as labeled.
}
  \label{fig:4}
\end{figure}
properties. Note that the solid curve for $N(\omega)/N(0)$
which is based on an MMP form with $\omega_{SF}=41\,$meV shows
no distinct sharp structure. Thus, a relatively smooth spectrum
extending over a large energy scale produces in the quasiparticle
density of states only very small, gradual modulations which
would be hard but not impossible to detect.

\section{Numerical Results}
\label{sec:4}

In Fig.~\ref{fig:4} we show our numerical results for the real
part of the quasiparticle self energy $\Sigma_1(\omega)$ 
(only for $\omega>0$ and made positive) as a
function of frequency $\omega$ based on numerical evaluation of
Eqs.~\eqref{eq:17} for the renormalized Matsubara frequencies. The
black dotted curves apply to the normal and the black solid lines
to the superconducting state. Both are at temperature $T=10\,$K.
The spectrum $I^2\chi(\omega)$ used in the top frame of Fig.~\ref{fig:4}
is the MMP form of Eq.~\eqref{eq:8} with $\omega_{SF}=41\,$meV
to which we have added a delta function contribution also at
$\omega_E=41\,$meV. This is shown as the dashed line in the top frame of
Fig.~\ref{fig:3}.
The resulting boson structure in the normal state self
energy is sharp and falls exactly at $\omega_E=41\,$meV. It is the
presence of the $\delta$-function in $I^2\chi(\omega)$ which makes
the peak so prominent. The corresponding peak in the superconducting
state (solid black curve) has shifted to higher frequency and falls
slightly below $\omega = \omega_E+\Delta_0 = 63.3\,$meV where
$\Delta_0$ is the gap amplitude. For an $s$-wave superconductor
the shift in the boson structure would fall exactly at
$\omega_E+\Delta_0$ but for $d$-wave it falls below this value
because the gap is distributed in value and we are seeing the
result of a distribution of shifts from 0 to $\Delta_0$. The
peak has also broadened and the weight under it appears to have
increased. In this sense, the opening up of the superconducting
gap in the quasiparticle density of states has not only shifted
the boson structure but has also, in a sense, enhanced it.

This same statement applies even more strikingly to the corresponding
optical quantities. For the imaginary part of the memory function,
optical mass renormalization $\lambda_{op}(\omega) = \{[(m^\ast(\omega)/m]%
-1\}$, black dashed curve for the normal state and black
dash-dotted curve for the superconducting state, the boson structure
is much greater in the dash-dotted curve. Comparing the normal state
memory function (black dashed curve) with the self energy (black dotted
curve) shows that the structure in the memory function is indeed at
$\sqrt{2}\omega_E$ rather than at $\omega_E$ although there is a
background to our boson spectrum besides the $\delta$-function.
This square root
of two shift becomes critical in serious comparison ARPES and
optical results and has not been appreciated in the past.\cite{hwang}
Note that for the superconducting case the structure in the memory function
is greatly enhanced over its normal state value and that it is also
shifted upwards in energy as for the self energy.
Its exact position
depends on the details of the $d$-wave gap structure and a complete
Eliashberg calculation is required to determine it. The size of the
peak in the underlying electron-boson spectral density $I^2\chi(\omega)$
is not related simply to the size of the structure seen in the
self energy or memory function.

To produce sharp, easily identifiable, structures in ARPES or optical
self energies, it is necessary to have correspondingly sharp
structures in $I^2\chi(\omega)$. This fact is well illustrated in
the bottom frame of Fig.~\ref{fig:4} where we show results based on
a simple MMP model, Eq.~\eqref{eq:18}, spectral density
$I^2\chi(\omega)$ (solid curve in the top frame of Fig.~\ref{fig:3}).
Direct comparison
of the curves in the bottom frame of Fig.~\ref{fig:4} with the
corresponding curves in the top frame shows that now the boson
structures are much reduced. Although there is a broad peak at
$41\,$meV in $I^2\chi(\omega)$ for the pure MMP spectrum, this does
not translate into a discernable peak in any of our results in the
bottom frame of Fig.~\ref{fig:4}. From this comparison we conclude
that a broad peak in $I^2\chi(\omega)$ is difficult to detect as a
clear signature in either the real part of the quasiparticle self
energy $\Sigma_1(\omega)$ or in the optical mass renormalization
$\omega\lambda_{op}(\omega)$ in the normal as well as in the
superconducting state. A detailed Eliashberg analysis
is needed to extract $I^2\chi(\omega)$ from such data.
A method for achieving this has been described in Refs.~\onlinecite{schach3}
and \onlinecite{schach4}. An appropriate derivative of
$\tau^{-1}_{op}(\omega)$, namely $d^2[\omega\tau^{-1}_{op}(\omega)]/%
d\omega^2$ which is closely related to $I^2\chi(\omega)$, see
Eq.~\eqref{eq:6}, is used to get a first modification to the
MMP form, Eq.~\eqref{eq:17}, that fits the optical data better. The
procedure can be continued until a reasonable fit with the optical
data is obtained from the model spectral density. However, no
high accuracy fit has so far been achieved.

In the absence of such an analysis experimentalists
have tried to get information on structure in $I^2\chi(\omega)$
through comparison between normal and superconducting state data.
To this end we show in the bottom frame of Fig.~\ref{fig:4}
results for $\omega\lambda_{op}(\omega)$ in the normal state at
$T=90\,$K (black dash-double dotted curve). A subtraction of
this data from the dash-dotted curve in the superconducting state gives
the difference (gray dash-dotted curve).
On comparing this curve with the input
$I^2\chi(\omega)$ (solid line, top frame of Fig.~\ref{fig:3})
we see little detail correspondence. The peak in the difference curve is not at
$41\,$meV and the MMP tails at larger values of $\omega$ are
not picked up. This holds as well if the normal state data at
$T=10\,$K (black dashed curve), albeit not accessible to
experiment, is used in the subtraction
(gray solid curve) or if $\Sigma_1(\omega)$ is considered
(gray dotted curve). This subtraction procedure does not give
a reliable way to relate boson structure in experimentally
measured quantities directly to boson structure in the
electron-boson spectral density and is not recommended as an
analysis procedure. This is also seen in the middle frame of
Fig.~\ref{fig:4} which applies to the case of the spectrum
fit to optical data on YBCO$_{6.95}$ twinned single crystals.\cite{schach3}
It is shown as the dotted curve in the top frame of Fig.~\ref{fig:3}.
It has a $\lambda=2.01$ and corresponds
to an MMP form modified with an optical resonance peak at
$\omega_r = 41\,$meV and there is zero weight at small $\omega$.
This spectrum is intermediate in ``sharpness of boson structure"
between the $\delta$-function and the pure MMP case. We see that
this reflected itself in how structured $\Sigma_1(\omega)$ and
$\omega\lambda_{op}(\omega)$ are. Note in particular that the large
peaks in the black dash-dotted curve for $\omega\lambda_{op}(\omega)$
is now shifted to higher frequencies in the middle frame when compared
to the top frame although the optical resonance peak is at the
same frequency as is the $\delta$-function in Fig.~\ref{fig:3}.
(This is because the gap $\Delta_0$ is bigger now.)
This demonstrates once more that no exact correspondence between
size and position of the structure in $\omega\lambda_{op}(\omega)$
and $I^2\chi(\omega)$ is possible without a detailed analysis,
particularly for extended spectra.
Also these functions strongly reflect the energy dependence of the
underlying quasiparticle density of states that exists in the
superconducting state (dotted curve, bottom frame of
Fig.~\ref{fig:3})
and it is not possible to, so to speak,
subtract out the effects of these modulations from the data.
Formally the subtraction procedure described above between
superconducting and normal state quantities when applied to the
middle frame of Fig.~\ref{fig:4} again fails to give reliable
information on the shape of the boson spectrum.

\begin{figure}[tp]
  \includegraphics[width=9cm]{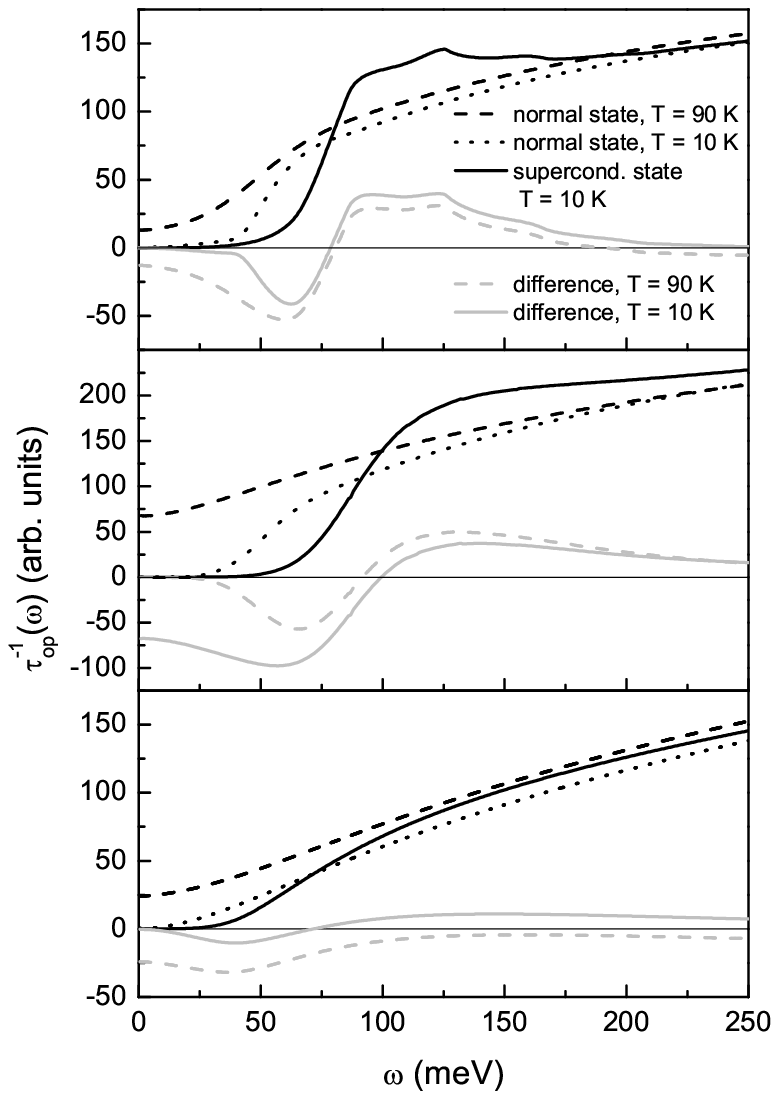}
  \caption{Optical scattering rates $\tau^{-1}_{op}(T,\omega)$
vs $\omega$ for the three quasiparticle-boson spectral
densities used in Fig.~\protect{\ref{fig:4}}. In each of the
three frames the solid black curve is for the superconducting state
at $T=10\,$K, the black dotted curve for the normal state at the
same temperature, and the black dashed curve for the normal state
at $T=90\,$K. The gray curves are the difference curves, solid
curve for $T=10\,$K and dashed curve for $T=90\,$K.
}
  \label{fig:5}
\end{figure}

In the top frame of Fig.~\ref{fig:5} we show results for the
optical scattering rate $\tau^{-1}_{op}(\omega)$ based on the
model $I^2\chi(\omega)$ which consists of an MMP form plus a
$\delta$-function (dashed curve in the top frame of Fig.~\ref{fig:3}) at
$\omega_E=41\,$meV. The solid black curve is in the superconducting
state at $T=10\,$K while the black dotted curve is in the normal
state at the same temperature. While the underlying $I^2\chi(\omega)$
is the same in both cases yet the amount of structure seen in the
solid curve is much more pronounced than it is for the dotted
curve. This is due entirely to the energy dependence of the
quasiparticle density of states (dashed line, bottom frame
of Fig.~\ref{fig:3})
which arises in a $d$-wave
superconductor. On comparing the solid and the dotted curve we
note a shift upward in the initial sharp rise in the scattering rate
as we go to the superconducting case. This corresponds to the
opening of the gap. The black dashed curve is in the normal state at
$T=90\,$K. Temperature smears the sharp rise at small $\omega$.
Only the $T=90\,$K data is available to experiment.
If, as we did in Fig.~\ref{fig:4}, we take the difference between
superconducting and normal state optical scattering rate we get the
gray solid line for $T=10\,$K and the gray dashed curve for
$T=90\,$K. The minimum in each of these curves is around $60\,$meV
which is $\omega_E$ shifted by approximately $\Delta_0$. It
is obviously hard to get reliable information on $\omega_E$
from such data.
The middle and bottom frames of Fig.~\ref{fig:5} are for less
structured $I^2\chi(\omega)$ and show a progressive decrease in
the corresponding structures in $\tau^{-1}_{op}(\omega)$.

We turn next to the exact correspondence between $\tau^{-1}_{op}(\omega)$
and the quasiparticle scattering rates given as twice
the imaginary part of the
self energy. In the previous section we have already examined this
relation but based our discussion on approximations for the
conductivity formula. Next we use the more exact Eq.~\eqref{eq:16}
and solutions of the full Eliashberg equations \eqref{eq:17}.

\section{Model Self Energy}
\label{sec:5}

We return to the model function $\tau^{-1}_{model}(\omega)$ defined in
Sec.~\ref{sec:3}. It can be constructed from the numerical data shown in
Fig.~\ref{fig:5} for $\tau^{-1}_{op}(\omega)$ defined as
\begin{equation}
  \label{eq:19}
  \tau^{-1}_{op}(T,\omega) = \frac{\Omega_p^2}{4\pi}\,
  {\rm Re}\,\sigma^{-1}(T,\omega) =
  \frac{\Omega_p^2}{4\pi}\frac{\sigma_1(T,\omega)}
  {\sigma_1^2(T,\omega)+\sigma_2^2(T,\omega)},
\end{equation}
and
\begin{equation}
  \label{eq:20}
  \omega\left(\frac{m^\ast}{m}\right)_{op} =
  \frac{\Omega_p^2}{4\pi}\,{\rm Im}\,\sigma^{-1}(T,\omega) =
  \frac{\Omega_p^2}{4\pi}\frac{\sigma_2(T,\omega)}
  {\sigma_1^2(T,\omega)+\sigma_2^2(T,\omega)}.
\end{equation}
We note in passing, that for small $\omega$ the optical effective
mass formula \eqref{eq:20} is dominated
by the imaginary part of the optical conductivity $\sigma_2(T,\omega)$
in the superconducting state because it diverges as $\omega^{-1}$ for small
$\omega$ with coefficients proportional to the inverse square of the
penetration depth. This is not related to a quasiparticle
effective mass, but is rather a property of the condensate itself.
Nevertheless, this is what has been done traditionally and we need
to follow this procedure here to make contact with the literature.
An alternate approach would be to subtract out of the imaginary part
of the optical conductivity the condensate contribution before
forming the ratios indicated in Eqs.~\eqref{eq:19} and \eqref{eq:20}.
In this alternate approach
the resulting $\sigma_1$ and $\sigma_2$ would refer directly only
to the normal fluid part of the optical conductivity and so would
be more closely related to quasiparticle properties. But this is
not what is done in the literature.

For comparison with the quasiparticle self energy it is useful
to use $\tau^{-1}_{model}(T,\omega)$ given by
\begin{equation}
  \label{eq:21}
   \tau^{-1}_{model}(T,\omega) = \frac{\Omega_p^2}{4\pi}
\frac{d}{d\omega}\left[\frac{\omega\sigma_1(T,\omega)}
  {\sigma_1^2(T,\omega)+\sigma_2^2(T,\omega)}\right]
\end{equation}
 than $\tau^{-1}_{op}(T,\omega)$. Its imaginary part is given by
the KK-transform
\begin{equation}
  \label{eq:22}
  \alpha(\omega) = -\frac{1}{\pi}\int_{-\infty}^\infty\!
  d\omega'\,\frac{\tau^{-1}_{model}(T,\omega')}{\omega'-\omega}.
\end{equation}

\begin{figure}[tp]
  \includegraphics[width=9cm]{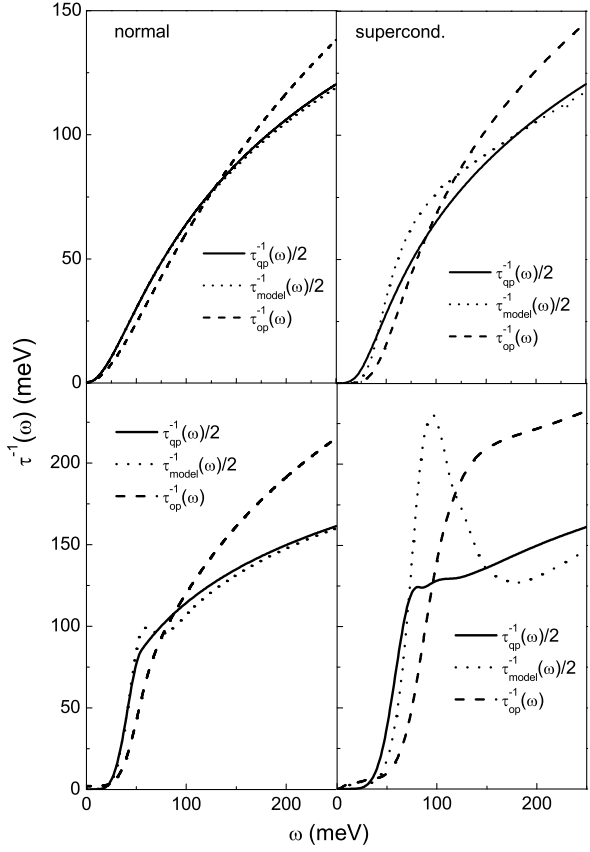}
  \caption{Comparison of optical, $\tau^{-1}_{op}(\omega)$ (dashed lines),
quasiparticle, $\tau^{-1}_{qp}(\omega)$ (solid lines), and model
$\tau^{-1}_{model}(\omega)$ (dotted lines) scattering rates vs $\omega$
for a temperature of $T=10\,$K. For
easier comparison $\tau^{-1}_{qp}(\omega)/2$ and
$\tau^{-1}_{model}(\omega)/2$ are shown. The two top frames compare
our results for an MMP form with $\omega_{SF}=41\,$meV in the
normal (left hand frame) and the superconducting state (right
hand frame). The two bottom frames give the results for
the model $I^2\chi(\omega)$
determined from optics for YBCO$_{6.95}$ as shown in the top
frame of Fig.~\protect{\ref{fig:3}} (dotted curve).
}
  \label{fig:6}
\end{figure}

In Fig.~\ref{fig:6} we show results for $\tau^{-1}_{model}(\omega)$
based on the numerical results of Fig.~\ref{fig:5} and compare
directly with the imaginary part of the self energy $\Sigma_2(\omega)$
obtained directly from Eliashberg calculations, namely
${\rm Im}\Sigma(\omega) = -{\rm Im}\tilde{\omega}(\omega+i0^+)$
of Eq.~\eqref{eq:17}. By definition of the self energy
$\tilde{\omega}(\omega+i0^+) = \omega-\Sigma(\omega+i0^+)$. In
the two top frames of Fig.~\ref{fig:6} we show results for the MMP form
(solid line in the top frame of Fig.~\ref{fig:3}) and in the
two bottom frames the
YBCO$_{6.95}$ form (dotted curve in the top frame of Fig.~\ref{fig:3}) for
$I^2\chi(\omega)$ is used. In each case we present two frames, the
left hand frame is for the normal state and the other frame for the
superconducting state.
All results are at temperature $T=10\,$K. For ease of comparison with
$\tau^{-1}_{op}(\omega)$ (dashed lines) it is $\tau^{-1}_{qp}(\omega)/2$
(solid lines) and
$\tau^{-1}_{model}(\omega)/2$ (dotted lines) that is
shown. Even when a factor of one half is included to make magnitude more
comparable, the optical scattering rates
differ significantly from the quasiparticle
scattering rates. On the
other hand, in the normal state the model scattering rate
agrees almost perfectly with the quasiparticle scattering
rate (top left hand frame).
It is clear that $\tau^{-1}_{model}(\omega)$ should be
used in comparison with ARPES data and not $\tau^{-1}_{op}(\omega)$.
The close correspondence between model and optical rates
is, however, lost when the superconducting state is considered
(top right hand frame). The
solid curve is significantly different from the dotted
curve. In particular, $\tau^{-1}_{model}(\omega)$ is smaller at
small $\omega$ and rises faster around $50\,$meV after which it
stays significantly above its quasiparticle counterpart up to
almost $175\,$meV where the two curves cross again. These differences
have their origin in the energy dependence of the superconducting
quasiparticle density of states (solid line, bottom frame of
Fig.~\ref{fig:3}) and have nothing
to do with the boson structure. For a comparison of
$\tau^{-1}_{op}(\omega)$ (optics) and
$\tau^{-1}_{qp}(\omega)$ (ARPES) based completely on experiment, the
reader is refered to Kaminski {\it et al.}\cite{kaminski2}

When the underlying $I^2\chi(\omega)$ is more structured, as is the case
for the YBCO$_{6.95}$ spectrum (dotted curve in Fig.~\ref{fig:3}),
the correspondence between $\tau^{-1}_{model}(\omega)$ and
$\tau^{-1}_{qp}(\omega)$ is not as good as is seen in the two
bottom frames of
Fig.~\ref{fig:6}. There is a small additional oscillation in the
model case, not present in the self energy.
Nevertheless, the agreement between them is still much closer than
is the case for the optical rate (dashed curve). Note,
however, that for the superconducting state the disagreement between
$\tau^{-1}_{qp}(\omega)$ (solid curve) and
$\tau^{-1}_{model}(\omega)$ (dotted curve) is now much greater
and, in particular, a large peak is seen in the model curve
which is not there in the dotted curve. This peak can be
related to the broad shoulder in the quasiparticle density of
states which follows the logarithmic singularity (dotted line, bottom
frame of Fig.~\ref{fig:3}).

\section{Conclusion}
\label{sec:6}

We have analyzed the relationship between the boson structure
seen in the quasiparticle self energy (be it real or imaginary part)
measured in ARPES experiments and the corresponding structure seen
in the optical memory function determined in the infrared
conductivity measurements. Starting first with the real part of
the self energy a $\delta$-function peak at $\omega=\omega_E$
in the quasiparticle-boson spectral density $I^2\chi(\omega)$ shows
up as a logarithmic-like peak at $\omega=\omega_E$
in the normal state. In the superconducting state
a gap develops in the quasiparticle density of states and this
introduces further structures in quasiparticle quantities. This effectively
shifts to higher energy the normal state boson structure, broadens it,
and can make it appear more prominent. For an $s$-wave gap the
shift would be $\Delta_0$ (gap amplitude) but for $d$-wave it is
somewhat less
because a distribution of gap values is involved. By contrast,
for the memory function the boson structure in the optical mass
renormalization is at $\sqrt{2}\omega_E$
in the normal state rather than at $\omega_E$ and it is much less
prominent. Only a very small peak results which greatly reduces
the value of such measurements for determining boson structure
as was attempted recently in Ref.~\onlinecite{hwang}. Again,
the boson structure shifts in the superconducting state and can also
appear more prominent as a result of the additional modulation
brought about by the opening of a gap.
These modifications, however, have nothing to do with the structures in
$I^2\chi(\omega)$. The two effects are not additive
and require a full non linear Eliashberg analysis to disentangle in detail
as we have provided here. When extended rather than $\delta$-function
spectra are used, the situation is even more complex. For example,
a rather broad peak in $I^2\chi(\omega)$, as in the MMP form for
spin fluctuations produces no peak at all, even in the superconducting
state.

So far we have described only the real part of the self energy
and equivalent optical quantity. Much the same can be said about
scattering rates. For a $\delta$-function $I^2\chi(\omega)$, the
quasiparticle scattering rate jumps from zero to a finite value
at $\omega=\omega_E$ and remains unchanged after that. On the
other hand, the optical scattering rate starts from zero at
$\omega=\omega_E$ and increases gradually towards the same finite
value which it only attains at very high $\omega$. Thus, the
signature of a $\delta$-function is not singular as it is for
the quasiparticle case.

We have also
found that in the normal state a new model scattering rate can
be introduced which we denote $\tau^{-1}_{model}(T,\omega)$
and define as $d[\omega\tau^{-1}_{op}(T,\omega)]/d\omega$.
It is completely determined from optical data and has the
advantage that it follows much more closely the $\omega$
dependence of the quasiparticle scattering rate than does
$\tau^{-1}_{op}(T,\omega)$ itself. Its KK-transform is very close
to the real part of the quasiparticle self energy. We propose
that it is this quantity that should be used in comparison between
ARPES and optical experimental data. The comparison will be close
in isotropic systems and in the anisotropic case, the model quantity
gives an average over all directions in the Brillouin zone. In
the superconducting state no such simple comparison between ARPES
and optics is possible. This needs to be kept in mind when analyzing
experimental data, particularly in studies aimed at deriving boson
structure from such data. This fact does not seem to have always
been appreciated in analysis of experimental data, where features, at
least partially associated with the superconducting quasiparticle
density of states, has been assigned to boson structure.

The model spectral density $I^2\chi(\omega)$ consisting
of an MMP form with superimposed the $\delta$-peak was used
to demonstrate that optics as well as ARPES just pick up the
sharp structures and do not give direct information on a possible
coupling to a smooth background. Nevertheless, it may well be
that it is precisely this coupling of the quasiparticles to such a
background which is primarily responsible for superconductivity.

\section*{Acknowledgment}
 
Research supported by the Natural Sciences and Engineering
Research Council of Canada (NSERC) and by the Canadian
Institute for Advanced Research (CIAR).

\end{document}